\documentstyle[12pt,nato]{article}
\input{epsf}
\begin{document}

\title{RELATIVISTIC FRACTAL COSMOLOGIES}

\author{Marcelo B. Ribeiro \dag \footnote{ \ Present address: Physics
  Institute, Federal University of Rio de Janeiro - UFRJ, Brazil;
  e-mail: mbr@if.ufrj.br}}. 

\affil{\dag \ Observat\'orio Nacional, Rio de Janeiro, BRAZIL}

%%%%%%%%%%%%%%%%%%%%%%%%%%%%%%%%%%%%%%%%%%%%%%%%%%% abstract

\beginabstract
This article presents a review of an approach for constructing a simple
relativistic fractal cosmology, whose main aim is to model the observed
inhomogeneities of the distribution of galaxies by means of the Tolman
solution of Einstein's field equations for spherically symmetric dust
in comoving coordinates. Such model is based on earlier works developed
by L.\ Pietronero and J.\ R.\ Wertz on Newtonian cosmology, and the main
points of these models are also discussed. Observational relations in
Tolman's spacetime are presented, together with a strategy for finding
numerical solutions which approximate an averaged and smoothed out single
fractal structure in the past light cone. Such fractal solutions are
actually obtained and one of them is found to be in agreement with basic
observational constraints, namely the linearity of the redshift-distance
relation for $z < 1$, the decay of the average density with the distance as
a power law (the de Vaucouleurs' density power law), the fractal dimension
within the range $1 \le D \le 2$, and the present range of uncertainty
for the Hubble constant. The spatially homogeneous Friedmann model is
discussed as a special case of the Tolman solution, and it is found that
once we apply the observational relations developed for the fractal model
we find that all Friedmann models look inhomogeneous along the backward
null cone, with a departure from the observable homogeneous region at
relatively close ranges. It is also shown that with these same observational
relations the Einstein-de Sitter model can have an interpretation where
it has zero global density, a result consistent with the ``zero global
density postulate'' advanced by Wertz for hierarchical cosmologies and
conjectured by Pietronero for fractal cosmological models. The article
ends with a brief discussion on the possible link between this model and
nonlinear and chaotic dynamics. 
\endabstract

%%%%%%%%%%%%%%%%%%%%%%%%%%%%%%%%%%%%%%%%%%%%%%%%%%% sections

\section{INTRODUCTION}

\indent

In the study of dynamical systems it is usual to approach the problem
under consideration by following a more or less pre-determined path,
which can be sketched as being basically made of four main stages.
In the {\sc first} step one usually sets up the differential or
difference equations
that describe the system in study, that is, one defines the dynamical
system itself. The {\sc second} step would be the determination of
orbits of the dynamical system, their discretization, that is,
determining the Poincar\'{e} section, etc, while the {\sc third} stage
usually consists of attempting to determine whether or not the system
exhibits sensitivity to initial conditions, and this could be done by
studying the Lyapunov exponents of the orbits. In other words, at this
stage one usually seeks to determine whether or not the dynamical
system is chaotic. Then the {\sc fourth} stage would be the search for
fractal self-similarity and strange attractors, with the possible
determination of the fractal dimension of the self-similar pattern.

I will call this way of studying dynamical systems as ``the
mathematicians' approach'' as it is usually followed by them, and the
point I would like to emphasize is that in such an approach the
self-similarity exhibited by the system, and the determination of its
fractal dimension, will play a minor and side role, and may even be
expendable in the whole treatment since when one finally reaches that
stage most of the dynamical behaviour of the system would already be
elucidated.

However, when a physicist looks at his or her noisy data, or in a
cosmological context, at his or her messy observations, the physicist
will usually attempt to get some sense from his/her real data, and in
this process he or she may ask: is this noisy data really noisy? Could
some sort of pattern be identified in this data? Then, the physicist may
go even further and pose the following question. Could a chaotic
behaviour be somehow represented in this data?

To attempt to answer those questions, in particular the last one,
the physicist might wish to approach the problem from the opposite
direction taken by the mathematicians, and a path that could be taken
may be sketched as follows. The {\sc first} stage would be to use the
self-similarity exhibited by the system in order to determine its fractal
dimension, together with some minimal and basic dynamical assumptions about
the system such that one starts with a workable and testable model.
Then the {\sc second} step would be to try to determine orbits, compute
the Lyapunov exponents, make the Poincar\'{e} map, try to find the attractors,
etc, of the system in order to reach the {\sc third} stage which would be
to ascertain whether or not the system exhibits chaos. The {\sc fourth}
stage would consist in the determination of the physical processes and
their observable effects that such possible chaotic behaviour would
bring to the system under study.

It is therefore clear that in this ``physicist's approach'' fractals would
play an essential and primary role in modelling natural phenomena as the
physicist would start by attempting to model structures we can see, assuming
implicitly what is in essence Maldelbrot's view of fractal geometry
\cite{mandelbrot} where one seeks to model naturally occurring shapes like
mountains, clouds and coastlines. However, in order to follow the path outlined
above, the physicist needs first to do two things:
\begin{enumerate}
  \item recognize a fractal pattern in the data;
  \item characterize it in some way.
\end{enumerate}
Unfortunately, these two points lead to questions that turn out to be
easy to ask but hard to answer, and in a cosmological context both
points above are currently controversial, especially the first one,
inasmuch as those who voiced their recognition of a fractal
pattern in the large-scale distribution of galaxies have faced stiff
resistance and, not rarely, hostility.

This article intends to present a review of a specific approach to the use of
fractals in relativistic cosmology inspired in Luciano Pietronero's Newtonian
model \cite{pietronero}, although most of the treatment was first
developed by James R.\ Wertz in his PhD thesis \cite{wertz}, where he studied
Newtonian Hierarchical Cosmology, and rediscovered by Pietronero by means
of the fractal language, as we shall see next. The underlying
philosophy of this approach is, as discussed above, to take advantage of
the ability of fractals in modelling shapes, and the basic aim is to
find solutions of Einstein's field equations with an approximate
{\it single} fractal behaviour along the past light cone. Therefore,
multifractals are not considered here. This is basically a
descriptive model which resembles Maldelbrot's characterization of
coastlines as fractal shapes \cite{mandelbrot}, and so far only covers
the first stage of the physicist's approach to the study of dynamical
systems as described above.

\section{MODERN COSMOLOGY AND FRACTALS}

\indent

The goal of modern cosmology is to find the large-scale matter
distribution and spacetime structure of the universe from astronomical
observations and, broadly speaking, there are two different ways of
approaching this problem \cite{ellis-stoeger}. The most popular and
standard approach consists of some {\it a priori} assumptions about the
geometry of the universe, usually based on pragmatic and philosophical
reasons. This generally reduces the cosmological problem to the
determination of a few free parameters that characterize such universe
models, and this determination then becomes the primary objective of
observational cosmology. That standard approach assumes the {\it
cosmological principle}, where the universe is thought to be spatially
homogeneous and isotropic on large-scales, and is represented by the
maximally symmetric Friedmann spacetime.

The alternative approach of studying cosmology is to attempt as far
as possible to determine spacetime geometry directly from astronomical
observations, where any kind of a priori assumptions are kept to a
minimum dictated by the essential and basic requirements necessary
such that one is able to construct a workable model.

The relativistic fractal cosmology of this article approaches the
cosmological problem from this alternative point of view, although it
must be said, it is obviously just one possible way of looking at
this problem. Considering recent astronomical observations as fundamental
empirical facts, I shall attempt to ``guess'' a metric in a specific
form suggested by those same observations which enables us to model a
more real, lumpy universe as detected in our past light cone.

The fundamental empirical facts of this approach to cosmology are the
recent all-sky redshift surveys
\cite{lapparent,kopylov,geller,geller-huchra,saunders,ramella} where it
is clear that the large-scale structure of the universe does not show
itself as a smooth and homogeneous distribution of luminous matter as
was thought earlier. Rather the opposite, since up to the limits of the
observations presented in those surveys, the three-dimensional cone maps
show a very inhomogeneous picture, with galaxies mainly grouped in
clusters or groups alongside regions devoid of galaxies, virtually empty
spaces with scales of the same order of magnitude as their neighbour
clusters. Such mapping of the skies gives us a picture of the
distribution of galaxies as a complex mixture of interconnected voids,
clusters and superclusters.

From these observations, what is obvious for the eyes is the pattern
that appears to be common in all surveys: the deeper the probing is
made, the more similar structures are observed and mapped, with clusters
turning into superclusters and even bigger voids being identified. As
the size of these clusters are only limited by the depth of the surveys
themselves, there is so far no empirical evidence of where and if those
structures finish.

With respect to this pattern, two ideas seem to fit in. The first is the
old concept of {\it hierarchical clustering} \cite{charlier1,charlier2}
which states that galaxies join together to form clusters that form
superclusters which themselves are elements of super-superclusters and
so on, possibly ad infinitum. The second is the more recent concept of
{\it fractals}, or self-similar structures, of which a rather loose tentative
definition proposed by Mandelbrot seems to be adequate for the present
purpose: ``a fractal is a shape made of parts similar to the whole in
some way'' (see \cite{feder}, p.\ 11).
In this context sef-similarity means that a fractal consists of a system
in which more and more structure appears at smaller and smaller scales
and the structure at small scales is similar to the one at large scales.
In other words, the same structure repeats itself at different scales.
It is therefore clear that fractals are a more precise version of the
same {\it scaling} idea behind the concept of hierarchical clustering, where
clusters and superclusters form a self-similar pattern repeating itself
at different scales. Consequently, in this context to talk about a
hierarchical structure is basically the same as to talk about a fractal
system.

The attempt of modelling the large-scale structure of the universe
as a hierarchical or fractal structure is not at all a new idea, and in
order to give some historical perspective of this problem I shall
present below a historical summary of the main contributions of such
attempts.~\footnote{ \ The history of the hierarchical clustering
hypothesis is of being forgotten by most every time it is voiced by some
as being a natural way of constructing the universe, only to be resurrected
by somebody else, just a while later, who not rarely is partially or
completely unaware of many of the previous results and studies. Therefore, it is
quite reasonable to believe that this historical summary may still be
revised once possible isolated and forgotten works are unearthed.}

\begin{description}

 \item[1907] E.\ E.\ Fournier D'Albe published the proposal of a
	hierarchical construction of the universe \cite{fournier};

 \item[1908,1922] C.\ V.\ L.\ Charlier applied Fournier D'Albe's idea
	to an astronomical world model in order to explain the Olber's
	paradox \cite{charlier1,charlier2};

 \item[1922-1924] F.\ Selety and A.\ Einstein discussed ``Charlier's
 	hierarchical model'' \cite{selety1,einstein,selety2,selety3}.

 \item[1970] G.\ de Vaucouleurs resurrected Charlier's model
 	\cite{vaucouleurs,vaucouleurs2} by proposing a hierarchical 
	cosmology as a way of explaining his finding that galaxies seem to 
        follow an average density power law with negative slope, of the form
        $\bar{\rho} \propto r^{-1.7}$; \linebreak J.\ R.\ Wertz studied some
	possible models of a Newtonian hierarchical cosmology
	\cite{wertz,wertz1}, and his simplified ``regular polka-dot model''
	anticipated Pietronero's \cite{pietronero} fractal treatment;

 \item[1971-1972] M.\ J.\ Haggerty and J.\ R.\ Wertz developed further
       the Newtonian hierarchical models and proposed some possible
       observational tests \cite{haggerty,haggerty-wertz}. The debate
       concerning the motivations \cite{vaucouleurs-wertz} and
       observational feasibility of hierarchical cosmological models is
       re-initiated and centered around the possible deviations from
       local expansion \cite{haggerty-wertz,sandage}. Such debate
       continued for over a decade or so, and now it seems to favour the
       view that there really are such deviations (see
       \cite{vaucouleurs3,feng} and references therein);

 \item[1972] W.\ B.\ Bonnor studied \cite{bonnor} a relativistic model
       in a Tolman spacetime with the de Vaucouleurs' density power law 
       $\bar{\rho} \propto r^{-1.7}$;

 \item[1978-1979] P.\ S.\ Wesson studied relativistic models
       \cite{wesson1,wesson2} with homothetic self-similarity where the
       local density followed $\rho \propto r^{-2}$ at some epochs;

 \item[1987] L.\ Pietronero published his fractal model
       \cite{pietronero} where the de Vaucouleurs density power law 
       $\bar{\rho} \propto r^{- \gamma}$ is obtained once one assumes that
       the large-scale distribution of galaxies forms a self-similar
       single fractal system. He also voiced sharp criticisms of
       the use of the spatial two-point correlation function for the
       characterization of the distribution of galaxies, in particular
       its indication that a homogenization of the distribution occurs at
       about 5 Mpc, a result he called ``spurious'' and due to the
       widespread unquestioned, and often unjustified, assumption of the
       untested homogeneous hypothesis. The debate still rages on
       \cite{coleman,davis,peebles,calzetti,coleman-pietronero,maurogordato};

 \item[1988] R.\ Ruffini, D.\ J.\ Song \& S.\ Taraglio proposed that the
       fractal system should have an upper cutoff to homogeneity in
       order to solve what they believed was an apparent conflict between
       ``the commonly accepted idea in theoretical cosmology that
       greater distances represent earlier epochs of the Universe
       implying that higher average densities should be observed'', and
       the de Vaucouleurs density power law that implies the opposite
       \cite{ruffini-song-taraglio};~\footnote{ \ Actually, such a
       possible transition to homogeneity had already been suggested
       by Wertz, \cite{wertz} p.~27.} D.\ Calzetti, M.\ Giavalisco \&
       R.\ Ruffini studied the spatial two-point correlation function
       for galaxies under a fractal perspective \cite{calzetti2} and
       reached some conclusions similar to Pietronero's \cite{pietronero}.

\end{description}

Since 1988 there has been a flurry of activity on fractal cosmology, and
the different approaches to the problem are in many ways opposing to
each other, a fact which has obviously created a lot of debate about the
issue. Nevertheless, from this historical summary it is clear that
although hierarchical cosmology has never belonged to the mainstream
of research in cosmology (at least until recently), it did attract the
attention of many researchers, who seriously considered the observed 
scaling behaviour of the distribution of galaxies as something fundamental
in cosmology.

Despite such interest, the most recent pre-fractal hierarchical
cosmological models suffered from some important shortcomings which may
explain why their development was hampered. In the first place it was
not generally clear what one meant mathematically by the idea of
``galaxies joining to form clusters that form superclusters which
themselves are elements of superclusters, and so on'', and such lack of
clarity had the effect of making the hierarchical concept not well defined
mathematically.~\footnote{ \ It should be noted here that Wertz
\cite{wertz,wertz1} did discuss the hierarchical concept in greater
mathematical details, but that work seems to have attracted little
attention since it has been published.} From a fractal perspective,
it is obvious that what is behind the hierarchical clustering concept
is the scaling behaviour empirically accepted by many in the
distribution of galaxies, and hence the main descriptor of such
behaviour becomes the fractal dimension which must be appropriately
defined in the context of the distribution of galaxies. Therefore, it
is the exponent of the de Vaucouleurs density power
law~\footnote{ \ From now on I shall call by ``de Vaucouleurs density
power law'' the decay of the average density of the distribution of
galaxies with the distance (using whatever definition of distance),
and given generically by the form $ \bar{\rho} \propto r^{- \gamma}$, where
$0 \le \gamma < 3$, that is, $\gamma$ does not necessarily have to be the
specific value of 1.7 as originally found by de Vaucouleurs
\cite{vaucouleurs}.} that has a fundamental physical meaning, and not
the law itself which is a consequence of the self-similarity of the
distribution of galaxies \cite{pietronero2}. However, the hierarchical
clustering concept had to wait until the appearance of fractals and the
works of Mandelbrot \cite{mandelbrot} and Pietronero \cite{pietronero},
who clarified this point.

A second difficulty of pre-fractal hierarchical cosmology was their
relativistic models which have never expressed distance by means of
observables, always using the unobservable coordinate distance $r$ (in
the spherically symmetric models) as measure of distance. That created
an additional difficulty between the empirically observed hierarchical
concept and the relativistic models, as the latter became in fact distorted
attempts of modelling the de Vaucouleurs' density power law since the
relativistic hierarchy was unobservable. Wesson \cite{wesson1,wesson2}
even gave up using the average density $\bar{\rho}$ and replaced it by
the local density $\rho$ in his version of de Vaucouleur's density power
law, and this inevitably led him to produce a model where the essence of
the original hierarchical clustering hypothesis was lost. On top of
all that, none of the relativistic models evaluated densities along
the past null geodesic despite the fact that General Relativity states
this is where astronomical observations are actually made. Consequently,
the pre-fractal relativistic hierarchical cosmologies were in fact ill
represented attempts of modelling hierarchical clustering, caused
mainly by the usual difficulties of modelling physical phenomena by
means of General Relativity conjugated with the additional difficulties
of solving the past null geodesic equation. Nevertheless, despite their
shortcomings those initial models did put forward interesting ideas
and methods, and as we shall see next the present approach to
relativistic fractal cosmology borrows some ideas from those
previous relativistic models and attempts to address and overcome most
of these problems, with the aim of recapturing the empirical and scaling
essence of the de Vaucouleurs density power law.

\section{PIETRONERO AND WERTZ'S MODELS}

\indent

Before we discuss the relativistic fractal models themselves, let us
briefly see the main most basic points of both Pietronero and Wertz's
models and how they compare to each other. While the former model is nowadays
relatively well-known in the literature, the latter seems to have attracted
very little attention since its publication as it is very rarely quoted.
For this reason I shall produce a more detailed presentation of the basis
of Wertz's model such that one may be able to fully appreciate its importance.

\begin{figure}[bht]
  \vspace{9cm}
  \epsfbox[0 0 30 50]{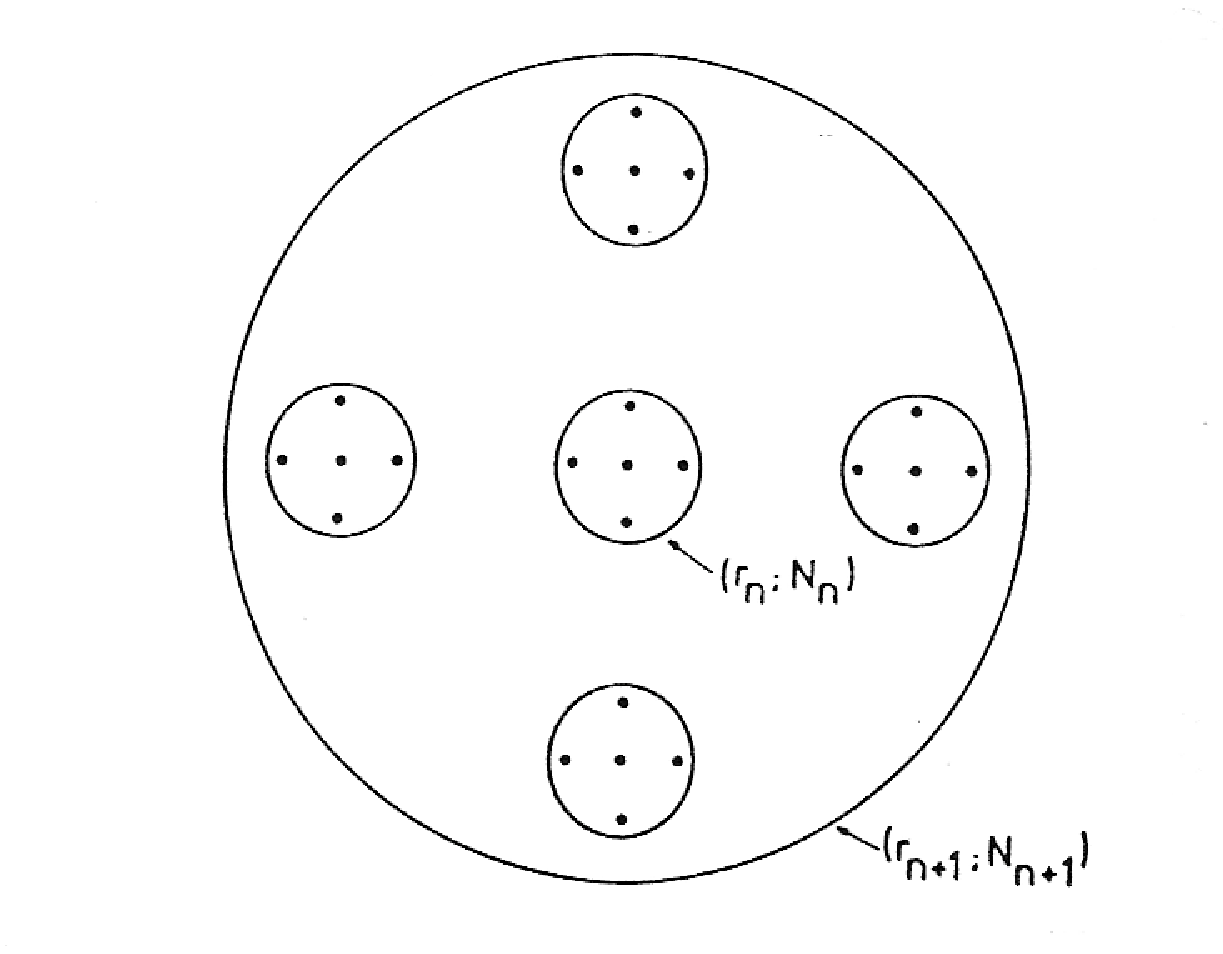}
  \caption{Reproduction from \protect\cite{pietronero} of a schematic
	   illustration of a deterministic fractal system from where a
	   fractal dimension can be derived. The structure is
	   self-similar, repeating itself at different
	   scales.}\label{figura-pietronero}
\end{figure}

In developing his model, Pietronero \cite{pietronero} started with a schematic
representation of a deterministic fractal structure, reproduced
in figure \ref{figura-pietronero}, whose basic idea is as follows. If
we start from a point occupied by an object and count how many objects
are present within a volume characterized by a certain length scale, we
get $N_0$ objects within a radius $r_0$, $N_1= \tilde{k} N_0$ objects within a
radius $r_1=k r_0$, $N_2=\tilde{k}N_1={\tilde{k}}^2 N_0$ objects within
a radius $r_2=kr_1=k^2r_0$. In general we have
\begin{equation}
  N_n={\tilde{k}}^n N_0
  \label{e1}
\end{equation}
within
\begin{equation}
  r_n=k^n \ r_0,
  \label{e2}
\end{equation}
where $\tilde{k}$ and $k$ are constants. By taking the logarithm of
equations (\ref{e1}) and (\ref{e2}) and dividing one by the other we get 
\begin{equation}
  N_n = \sigma { \left( r_n \right) }^D,
  \label{e3}
\end{equation}
with
\begin{equation}
  \sigma \equiv \frac{N_0}{ { \left( {r_0} \right) }^D },
  \label{e3+1}
\end{equation}
\begin{equation}
  D \equiv \frac{ \log \tilde{k}}{ \log k},
  \label{e3+2}
\end{equation}
where $\sigma$ is a prefactor of proportionality related to the lower
cutoffs $N_0$ and $r_0$ of the fractal system, that is, the inner limit
where the fractal systems ends, and $D$ is the fractal dimension. If we
smooth out the fractal structure we get the continuum limit of equation
(\ref{e3}), 
\begin{equation}
  N(r)= \sigma r^D,
  \label{e4}
\end{equation}
called ``generalized mass-length relation'' by Pietronero.
The de Vaucouleurs density power law is obtained if we now suppose that
a portion of the fractal system is contained inside a spherical sample
of radius $R_s$. Then
\begin{equation}
 \langle \rho \rangle \equiv \frac{N(R_s)}{V(R_s)} =
 \frac{3 \sigma}{4 \pi} {(R_s)}^{- \gamma}; \ \ \ \gamma \equiv 3-D,
 \label{e5}
\end{equation}
where $\langle \rho \rangle$ is the average density of the distribution. If
we take the value of $\gamma$ as the one found by de Vaucouleurs
\cite{vaucouleurs}, this implies that the fractal dimension of the
distribution is $D \approx 1.3$.

These are the results of interest for the relativistic approach of
this article, and therefore I shall stop here this brief summary of
Pietronero's fractal model. The interested reader can find in
\cite{coleman-pietronero} a comprehensive account of this fractal
approach to cosmology, plus the controversy surrounding the spatial and
angular two-point correlation functions and a discussion on multifractals
in this context.

\begin{figure}[bht]
  \vspace{12.5cm}
  \epsfbox[0 0 30 50]{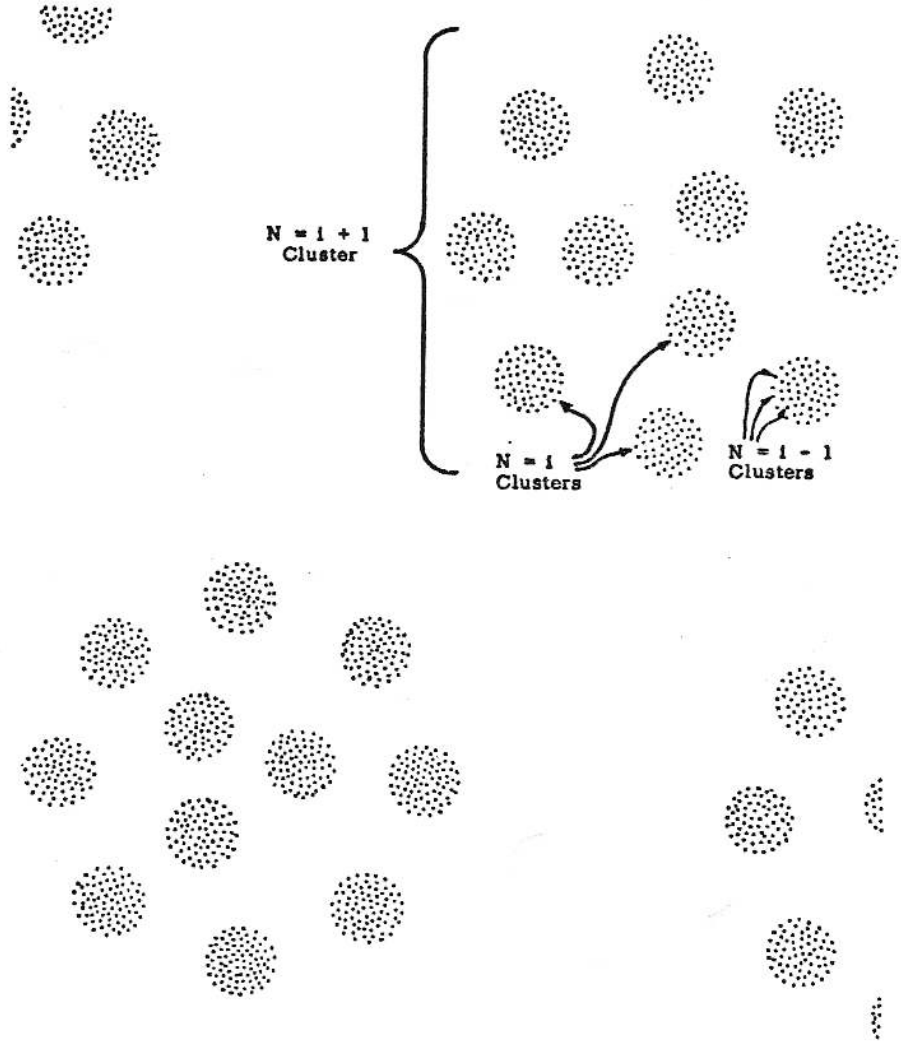}
  \caption{Reproduction from p.\ 25 of \protect\cite{wertz} of a rough
	   sketch cross-section of a portion of an $N=i+2$ cluster of a
	   polka dot model.}\label{figura-wertz}
\end{figure}

The hierarchical model advanced by Wertz \cite{wertz} was conceived at a
time when fractal ideas had not yet appeared, so in developing his
model Wertz was forced to start with a more conceptual discussion in order to
offer ``a clarification of what is meant by the `undefined notions' which
are the basis of any theory'' (\cite{wertz}, p.\ 3). Then he stated that
``a cluster consists of an aggregate or gathering of {\sc elements} into a
more or less well-defined group which can be to some extent distinguished
from its surroundings by its greater density of elements. A {\sc hierarchical}
structure exists when {\sc {\rm i}th order clusters} are themselves elements
of an {\sc {\rm (i+1)}th order cluster}. Thus, galaxies ({\sc zeroth order
clusters}) are grouped into {\sc first order cluster}. First order clusters
are themselves grouped together to form {\sc second order clusters}, etc,
{\it ad infinitum}'' (see figure \ref{figura-wertz}).

Although this sort of discussion may be very well to start with, it
demands a precise definition of what one means by a cluster in order to
put those ideas on a more solid footing, otherwise the hierarchical
structure one is talking
about continues to be a somewhat vague notion. Wertz seemed to have realized
this difficulty when later he added that ``to say what percentage
of galaxies occur in clusters is beyond the abilities of current
observations and involves the rather arbitrary judgment of what sort of
grouping is to be called a cluster. (...) It should be pointed out that
there is not a clear delineation between clusters and superclusters''~(p.\ 8).

Despite this initially descriptive and somewhat vague discussion about
hierarchical structure,
which is basically a discussion about scaling in the fractal sense,
Wertz did develop some more precise notions when he began to discuss
specific models for hierarchy, and his starting point was to assume what
he called the ``universal density-radius relation'', that is, the de
Vaucouleurs density power law, as a fundamental empirical fact to be
taken into account in order to develop a hierarchical cosmology. Then if
$M(x,r)$ is the total mass within a sphere of radius $r$ centered on the
point $x$, he defined the {\it volume density} $\rho_v$ as being the
average over a sphere of a given volume containing $M$. Thus
\begin{equation}
 \rho_v (x,r) \equiv \frac{3 M(x,r)}{4 \pi r^3},
 \label{e6}
\end{equation}
and the {\it global density} was defined as being 
\begin{equation}
 \rho_g \equiv \lim_{r \rightarrow \infty} \rho_v (x,r).
 \label{e7}
\end{equation}
A {\it pure hierarchy} is defined % by Wertz, \cite{wertz} p.\ 18,
as a model universe which meets the following pos\-tu\-lates:
\linebreak[4] {\it (i)}
for any positive value of $r$ in a bounded region, the volume density
has a maximum; \linebreak[4] {\it (ii)} the model is composed of only
mass points with finite non-zero mean mass; \linebreak[4] {\it (iii)}
the zero global density postulate: ``for a pure hierarchy the global
density exists and is zero everywhere'' (see \cite{wertz} p.\ 18).

With this picture in mind, Wertz states that ``in any model which involves
clustering, there may or may not appear discrete lengths which represent
clustering on different scales. If no such scales exist, one would have an
{\sc indefinite hierarchy} in which clusters of every size were equally
represented (...). At the other extreme is the {\sc discrete hierarchy} in
which cluster sizes form a discrete spectrum and the elements of one size
cluster are all clusters of the next lowest size'' (p.\ 23). Then in order 
to describe {\it polka dot models}, that is, structures in a discrete hierarchy
where the elements of a cluster are all of the same mass and are distributed
regularly in the sense of crystal lattice points, it becomes necessary for
one be able to assign some average properties. So if $N$ is the order of
a cluster, $N=i$ is a cluster of arbitrary order (figure~\ref{figura-wertz}),
and at least in terms of averages a cluster of mass $M_i$, diameter $D_i$ and
composed of $n_i$ elements, each of mass $m_i$ and diameter $d_i$, has
a density given by
\begin{equation}
 \rho_i = \frac{6 M_i}{\pi {D_i}^3}.
 \label{e8}
\end{equation}
From the definitions of discrete hierarchy it is obvious that
\begin{equation}
 M_{i-1} = n_{i-1} m_{i-1} = m_i,
 \label{e9}
\end{equation}
and if the ratio of radii of clusters is
\begin{equation}
 a_i \equiv \frac{D_i}{d_i} = \frac{D_i}{D_{i-1}},
 \label{e10}
\end{equation}
then the {\it dilution factor} is defined as 
\begin{equation}
 \phi_i \equiv \frac{\rho_{i-1}}{\rho_i} = \frac{{a_i}^3}{n_i} > 1,
 \label{e11}
\end{equation}
and the {\it thinning rate} is given by
\begin{equation}
 \theta_i \equiv \frac{\log (\rho_{i-1} / \rho_i)}{\log (D_i / D_{i-1})}
          =  \frac{ \log ({a_i}^3 / {n_i} ) }{\log a_i}.
 \label{e12}
\end{equation}

A {\it regular polka dot model} is defined as the one whose number of
elements per cluster $n_i$ and the ratio of the radii of successive clusters
$a_i$ are both constants and independent of $i$, that is, $n$ and $a$
respectively. Consequently, the dilution factor and the thinning rate are
both constants in those models,
\begin{equation}
  \phi=\frac{a^3}{n}, \ \ \ \ \theta=\frac{\log (a^3 /n)}{\log a}.
 \label{e13}
\end{equation}

The {\it continuous representation} of the regular polka dot model, which
amounts essentially to writing the hierarchical model as a continuous
distribution, is obtained if we consider $r$, the radius of spheres
centered on the origin, as a continuous variable. Then, from equation
(\ref{e10}) the radius of the elementary point mass $r_0$, is given by 
\begin{equation}
 r_0 = \frac{R_1}{a},
 \label{e14}
\end{equation}
where $R_N$ is the radius of a Nth order cluster with $M_N$ mass,
$V_N$ volume, and obviously that $R_0=r_0$. It follows from equation
(\ref{e14}) the relationship between $N$ and $r$,
\begin{equation}
 r= a^N r_0,
 \label{e15}
\end{equation}
where $R_N=r$. Notice that by doing this continuous representation Wertz 
ended up obtaining an equation (eq.\ \ref{e15}) which is nothing more
than exactly equation (\ref{e2}) of Pietronero's single fractal model,
although Wertz had reached it by means of a more convoluted reasoning.
Actually, the critical assumption which makes his polka dot model
essentially the same as Pietronero's fractal model was to assume the
regularity of the model because then $a$ and $n$ become constants. Also
notice that this continuous representation amounts to changing from 
discrete to an indefinite hierarchy, where in the latter the characteristic
length scales for clustering are absent. Therefore, in this
representation clusters (and voids) extend to all ranges where the
hierarchy is defined with their sizes extending to all scales between
the inner and possible outer limits of the hierarchy. Hence, in this
sense the continuous representation of the regular polka dot model has
exactly the same sort of properties as the fractal model discussed by
Pietronero.

From equation (\ref{e9}) we clearly get
\begin{equation}
 M_N = n^N M_0,
 \label{e16}
\end{equation}
which is equal to equation (\ref{e1}), except for a different notation,
and hence the de Vaucouleurs density power law is easily obtained as
\begin{equation}
 \rho_v = \frac{M_N}{V_N} = \left[ \frac{3 M_0}{4 \pi {r_0}^{ ( \log n /
           \log a ) }} \right] r^{- \theta },
 \label{e17}
\end{equation}
where $\theta$ is the thinning rate
\begin{equation}
   \theta =  3 - \left( \frac{ \log n}{ \log a} \right).
   \label{e18}
\end{equation}
Notice that equations (\ref{e17}) and (\ref{e18}) are exactly equations
(\ref{e5}), where $\gamma$ is now called the thinning rate. Finally, the
{\it differential density}, called {\it conditional density} by Pietronero, 
is defined as
\begin{equation}
 \rho_d \equiv \frac{1}{4 \pi r^2} \frac{ d M(r)}{d r} = 
	\left( 1 - \frac{\theta}{3} \right) \rho_v.
 \label{e19}
\end{equation}

From the presentation above it is then clear that from a geometrical
viewpoint Wertz's continuous representation of the regular polka dot
model is nothing more than Pietronero's single fractal model. However,
the two approaches may be distinguished from each other by some
important conceptual differences. Basically, as Pietronero clearly
defines the exponent of equation (\ref{e3}) as a fractal dimension, that
immediately links his model to the theory of critical phenomena in
physics, and also to nonlinear dynamical systems, bringing a completely
new perspective to the study of the distribution of galaxies, with
potentially new mathematical concepts and analytical tools to analyze
this problem. In addition, it also strongly emphasizes the fundamental
importance of scaling behaviour in the observed distribution of galaxies and the
exponent of the power law, as well as pointing out the appropriate
mathematical tool to describe this distribution, namely the fractal
dimension. All that is missing in Wertz's approach, and his thinning
rate is just another parameter in his description of hierarchy, without
any special physical meaning attached to it. Therefore, in this sense
his contribution started and remained as an isolated work, forgotten
by most, and which could even be viewed simply as an ingenious way
of modelling Charlier's hierarchy, but nothing more.

Nonetheless, it should be said that this discussion must not be
viewed as a critique of Wertz's work, but simply as a realization of the
fact that at Wertz's time nonlinear dynamics and fractal geometry were
not as developed as at Pietronero's time, if developed at all, and
therefore Wertz could not have benefited from those ideas. Despite this
it is interesting to note that with less data and mathematical concepts
he was nevertheless able to go fairly far in discussing scaling behaviour
in the distribution of galaxies, developing a model to describe it in the
context of Newtonian cosmology, and even suggesting some possible ways of
investigating relativistic hierarchical cosmology.

\section{A RELATIVISTIC APPROACH TO HIERARCHICAL (FRACTAL) COSMOLOGY}

\indent

In this section I shall start to develop a relativistic cosmology based
on the ideas expressed by Pietronero and Wertz and discussed in the
previous section. The notation used will be the same as in Pietronero's 
approach with minor changes, but some quantities will be referred by the
names used by Wertz.

The first thing necessary for one to start a relativistic model is
obviously the choice of the appropriate metric for the problem under
consideration. In the case of a relativistic fractal cosmology, we
need an inhomogeneous metric so that it becomes possible to derive a
relativistic version of Pietronero's relation (\ref{e4}). [{\it Note added
in 2009-- Regarding this point, a different perspective is advanced  in
Refs.\ \cite{rib2001b,rib2001,rib2001c} where fractals can be accommodated 
within a spatially homogeneous metric.}] Following
the simple geometrical ideas outlined by both Wertz and Pietronero,
spherical symmetry seems to be reasonable enough to start with. By
means of a similar reasoning, a dust distribution of matter also seems
reasonable enough to start with, and in that way we have outlined some
requirements which are equivalent to making some strong simplifications
that seem enough for a simple exploratory relativistic fractal
cosmological model. Bearing this discussion in mind, the {\it Tolman
solution} suggests itself as it is the general solution of the
Einstein's field equations for spherically symmetric dust in comoving
coordinates \cite{tolman}. However, the Tolman solution is
spherically symmetric about {\it one}
point, and constructing a cosmological model with it means that we would
be giving up the Copernican principle which states that there are no
preferred points in the universe. Despite this difficulty, if we assume
that there could be an upper cutoff to homogeneity in our fractal
system, we can construct a model using a variation of the
Einstein-Straus geometry, also known as ``Swiss-cheese'' models, with an
interior solution provided by the Tolman metric surrounded by a
Friedmann spacetime. Such a model needs to solve the junction conditions
between the two metrics in order to achieve a smooth transition, and the
solution imposes strong restrictions to the mass inside the Tolman
region, namely that the gravitational mass inside must be the same as if
the whole spacetime were Friedmannian and the Tolman region were never
there \cite{paper1}. By means of such scheme we are able to satisfy the
Copernican principle in our relativistic fractal model, but for reasons
that will become clear later, such geometry is not really
mandatory.~\footnote{ \ See \protect\cite{paper1} for a more detailed
discussion of the role played by the Copernican and cosmological
principles in fractal cosmologies. It is interesting to mention that
Wertz \protect\cite{wertz} did suggest the Swiss-cheese model as a
possible way of modelling relativistic hierarchy, and he also made a
discussion about the cosmological principle in the context of
hierarchical cosmologies.}

In order to make use of Tolman's models as descriptors of observations,
it is necessary first of all to derive the observational relations in
this metric. I shall present next a brief summary of the Tolman spacetime
followed by its observational relations, but without demonstration. Full
details about the observational relations in this metric can be found in
\cite{paper1}.

The Tolman metric with $\Lambda=0$ and $c=G=1$ may be written as
\begin{equation}
 dS^2=dt^2 - \frac{{R'}^2}{f^2}dr^2 - R^2 d \Omega^2; \ \ \ r \ge 0,
      \ \ \ R(r,t) > 0,
 \label{e22}
\end{equation}
where
\[
 d \Omega^2 = d \theta^2 + \sin^2 \theta d \phi^2,
\]
and $f(r)$ is an arbitrary function. The Einstein's field equations for the
metric (\ref{e22}) reduces to a single equation,
\begin{equation}
 2 R {\dot{R}}^2 + 2 R (1-f^2)=F,
 \label{e23}
\end{equation}
where $F(r)$ is another arbitrary function. The proper density is given by
\begin{equation}
 8 \pi \rho = \frac{F'}{2 R' R^2},
 \label{e24}
\end{equation}
and the dot means $\partial/\partial t$ and the prime $\partial/\partial r$.

The solution of equation (\ref{e23}) has three distinct cases according as
$f^2=1$, $f^2 > 1$ and $f^2 < 1$, these cases being termed, respectively,
parabolic, hyperbolic and elliptic Tolman models. In the parabolic models
($f^2=1$) the solution of equation (\ref{e23}) is
\begin{equation}
 R=\frac{1}{2} {(9 F)}^{1/3} { (t+ \beta) }^{2/3},
 \label{e25}
\end{equation}
where $\beta(r)$ is a third arbitrary function.~\footnote{ \ Actually, one
of the three functions $f(r)$, $F(r)$, $\beta(r)$ can be removed by a coordinate
transformation, so there really are two arbitrary functions in the Tolman
solution.} In the hyperbolic models ($f^2 > 1$) the solution of equation
(\ref{e23}) may be written in terms of a parameter $\Theta$
\begin{equation}
 R=\frac{F (\cosh 2 \Theta -1)}{4 (f^2 -1)},
 \label{e26}
\end{equation}
\begin{equation}
 t + \beta = \frac{ F ( \sinh 2 \Theta - 2 \Theta )}{4 {(f^2 -1)}^{3/2}},
 \label{e27}
\end{equation}
and finally in the elliptic models ($f^2 < 1$) the solution of equation
(\ref{e23}) may be written as
\begin{equation}
 R = \frac{F (1- \cos 2 \Theta)}{ 4 \mid f^2 -1 \mid },
 \label{e28}
\end{equation}
where
\begin{equation}
 t + \beta = \frac{F ( 2 \Theta - \sin 2 \Theta)}{ 4
                  { \mid f^2 -1 \mid }^{3/2}}.
 \label{e29}
\end{equation}
The three classes of the Tolman solution are the equivalent of flat, open and
closed Friedmann models.

The observational relations in Tolman's spacetime necessary in this work
have been calculated in \cite{paper1}, and I shall only state here the
results obtained. The {\it luminosity distance} in the Tolman model is
given by
\begin{equation}
 d_\ell=R {(1+z)}^2.
 \label{e30}
\end{equation}
The {\it redshift}  may be written as
\begin{equation}
  1+z={(1-I)}^{-1},
  \label{e31}
\end{equation}
where the function $I(r)$ is the solution of the differential
equation~\footnote{ \ The expression for the redshift in the Tolman model was
obtained in collaboration with \linebreak M.\ A.\ H.\ MacCallum.}
\begin{equation}
  \frac{dI}{dr}= \frac{{\dot{R}}'}{f} (1-I).
  \label{e32}
\end{equation}
The number of sources which lie at radial coordinate distances less than $r$
as seen by the observer at $r=0$ and along the past light cone, that is, the
{\it cumulative number count}, is given by
\begin{equation}
  N_c (r)= \frac{1}{4 M_G} \int \frac{F'}{f} dr,
  \label{e33}
\end{equation}
where $M_G$ is the average galactic rest mass ( $\sim 10^{11} M_\odot$). The
volume $V$ of the sphere which contains the sources, and the volume (average)
density $\rho_v$, have the form
\begin{equation}
 V (r) \equiv \frac{4}{3} \pi {(d_\ell)}^3 = \frac{4}{3} \pi R^3 {(1+z)}^6,
 \label{e34}
\end{equation}
\begin{equation}
 \rho_v \equiv \frac{N_c(r) M_G}{V(r)} = \frac{3}{16 \pi R^3 {(1+z)}^6}
                                         \int \frac{F'}{f} dr.
 \label{e35}
\end{equation}
The relativistic version of Pietronero's generalized mass-length relation
used in this work is given by
\begin{equation}
  N_c = \sigma {(d_\ell)}^D,
  \label{e36}
\end{equation}
and if we substitute equations (\ref{e34}) and (\ref{e36}) into equation
(\ref{e35}) we get a relativistic version of the de Vaucouleurs density
power law
\begin{equation}
 \rho_v = \frac{3 \sigma M_G}{4 \pi} {(d_\ell)}^{- \gamma},
 \label{e37}
\end{equation}
where
\begin{equation}
 \gamma = 3-D
 \label{e38}
\end{equation}
is Wertz's thinning rate. Equation (\ref{e37}) gives the volume density
for an observed sphere of certain radius $d_\ell$ that contains a portion
of the fractal system. All observational relations above must be calculated
along the past light cone, so if we adopt the radius coordinate $r$ as the
parameter along the backward null cone, we can then write the radial null
geodesic of metric
(\ref{e22}) as
\begin{equation}
 \frac{dt}{dr} = - \frac{R'}{f}.
 \label{e39}
\end{equation}
Notice that along the past light cone $R=R[r,t(r)]$, where $t(r)$ is the
solution of equation (\ref{e39}), and this is the function to be used in
the relations above such that they are really calculated along the null
geodesic.

The next step in order to find fractal solutions in the Tolman model is
to take advantage of its own freedom, given by the arbitrariness of the
functions $f(r)$, $F(r)$ and $\beta(r)$, such that we are able to simulate
the desired distribution of dust by means of a method similar to the one
employed by Bonnor \cite{bonnor}, but in a more complex and 
realistic context than his. However, in order to do so we have
to solve the past radial null geodesic (\ref{e39}) and then the equation
for the redshift (\ref{e32}), an almost impossible task to be carried
out analytically due to the fact that the functions $R$, $R'$, ${\dot{R}}'$
form complex algebraic expressions \cite{paper1}, and so a numerical solution
becomes inevitable. Such an approach can be essentially described as follows.

For a fractal structure as described above, the de Vaucouleurs density power 
law holds. Hence equation (\ref{e37}) may be written as
\begin{equation}
 \log \rho_v = a_1 + a_2 \log d_\ell,
 \label{e40}
\end{equation}
where $a_1$ and $a_2$ are constants related to the lower cutoff of the fractal
pattern and its fractal dimension, 
\begin{equation}
 D = a_2 + 3,
 \label{e41}
\end{equation}
\begin{equation}
  \sigma = \frac{4 \pi}{3 M_G} \mbox{exp} (a_1).
 \label{e42}
\end{equation}
Generally speaking, the numerical algorithm for finding those fractal 
solutions can be summarized as follows:
\begin{enumerate}
 \item {\sc start} by choosing $f(r)$, $F(r)$, $\beta(r)$;
 \item solve numerically the two ordinary differential equations (\ref{e32})
       and (\ref{e39});
 \item evaluate the observational relations along the past light cone: $d_\ell$,
       $\rho_v$, $N_c$, $z$, $V$;
 \item fit a straight line with the points obtained by numerical integration,
       according to equation (\ref{e40});
 \item is the fitting linear and with negative slope?
 \item if the answer is no, then choose other functions $f(r)$, $F(r)$,
       $\beta(r)$ and go all over again; else {\sc stop}: a fractal
       solution was modelled and $D$ and $\sigma$ are easily found.
\end{enumerate}
The description above is very schematic since many other important details
like root finding algorithm for equations (\ref{e27}) and (\ref{e29}), 
initial conditions for the {\sc ode}'s, numerical
integrator, etc, have to be considered. A full discussion of the numerical
problems involved in this method can be found in \cite{paper1,paper3}. In
any case, from the description above it is obvious that if the distribution
of dust remains homogeneous throughout the past null geodesic, the volume 
density will not change and $\log \rho_v = \mbox{constant}$, and $D=3$. In
this case the plot $\log \rho_v$ vs. $\log d_\ell$ will be a straight
line with zero slope.

Before closing this section, a few words are necessary here in order to
explain why the luminosity distance $d_\ell$ was chosen as the measurement
of distance in this relativistic approach to fractal cosmology. As is well
known, in relativistic cosmology we do not have an unique way of measuring
the distance between source and observer since their separation depends on
circumstances. We can, for instance, make use of geometrically defined
distances like the proper radius, or observationally defined distances like
the luminosity distance or the observer area distance (also known as angular
diameter distance) in order to say that a certain object lies at a certain
distance from us. The circumstances which tell us which definition to use
can also be determined on observational grounds, and so if we only have at
our disposal the apparent magnitudes of galaxies we associate to each of
them the luminosity distance and use such measurement in our analysis. On
the other hand, if these apparent magnitudes are corrected by the redshift
of the sources, we can then associate the corrected luminosity distance,
which is the same as the observer area distance obtained if we have the
apparent size of the objects \cite{ellis71}, and, therefore, another kind
of distance measure is obtained. Any of these observational distances are
as valid as any other, as real as any other, with the choice being dictated
by the availability of data, the nature of the problem being treated and
its convenience, but they will only have the same value at $z \ll 1$,
varying sometimes widely for larger $z$ (see  \cite{mcvittie} for a
comparison of these distances in simple cosmological models). In this
work we are interested in observables because we seek to compare theory
with observations, and this means that geometrical distances are of no
interest here. Consequently, the approach of this  relativistic cosmological
model is different from others where unobservable coordinate distances
(differences between coordinates) and separations (integration of the line
element $dS$ over some previously defined surface) are taken as measure of
distance, and in order to develop a treatment coherent with the
observational approach of this problem we need to make a choice among the
observational distances based on the nature of the problem and the
observations available. It is the intention of this work to propose a
relativistic extension of Pietronero-Wertz model in order to describe the
observed scaling behaviour of the distribution of galaxies detected by the
recent all-sky redshift surveys, and in this area of research it is usual 
for observers to take the luminosity distance as their indicator of
distance (for example, \cite{saunders2} does use $d_\ell$ in its
statistical analysis of a sample of {\sc iras} galaxies). It seems therefore
perfectly reasonable to take  the luminosity distance as the most appropriate
definition of distance to use in the context of this work, because what is
sought is to mimic the current methodology followed by many observers in
this field, and to carry out a comparison between the theoretical
predictions of this relativistic cosmological model and the observational
results brought by the redshift surveys.

\section{HOW INHOMOGENEOUS IS A ``HOMOGENEOUS'' UNIVERSE?}

\indent

The observational relations presented in the previous section
were derived with the clear intention of developing a relativistic fractal
model in the sense of Pietronero and Wertz, but as we shall see in this
section, the application of these same observational relations to the
spatially homogeneous Friedmann spacetime bring to light some very
interesting and unexpected results. [{\it Note added in 2009-- See Refs.\
\cite{rib1995,rib2001b,rib2001,rib2001c,rib2005,airs07,rlr2008} for 
further developments of these ideas and more in depth discussions on the
results below.}]

In order to see how those results are achieved, let me say first of all that
the question which motivated the application of the observational
relations above to the Friedmann spacetime came from the realization that
fractal dimensions, in the sense of Pietronero and Wertz, were defined in
Euclidean spaces, and it is not clear beforehand whether equation (\ref{e36})
will give the value $D=3$ for the fractal dimension in the Friedmann metric 
when we are dealing with observables. One can see more clearly this point if
we remember that this metric is {\it spatially} homogeneous, that is, it has
constant local densities at constant time coordinates, and when we integrate
along the past light cone, going through hypersurfaces of $t$ constant with
each one having different values for the proper density, one may argue that
$D$ could depart from the value 3 even in a spatially homogeneous spacetime
\cite{paper1}. With this point in mind, we may go even further and ask
whether or not even the Friedmann metric could be compatible with a fractal
description of cosmology \cite{paper2}, even in the strongly simplified
relativistic version of Pietronero-Wertz model presented here. In order to 
try to answer those questions, it is convenient to start with the analytically
feasible Einstein-de Sitter model.

It is shown in the appendix that the Einstein-de Sitter model can be obtained
from Tolman's spacetime as a special case when
\begin{equation}
 f(r)=1, \ \ \ F(r)= \frac{8}{9} r^3, \ \ \ \beta(r)=\beta_0,
 \label{e43}
\end{equation}
where $\beta_0$ is a constant. In this case the two differential equations
(\ref{e32}) and (\ref{e39}) are easily integrated from $I=t=r=0$ to $I(r)$
and $t(r)$, respectively yielding
\begin{equation}
 1-I= { \left( \frac{3 {\beta_0}^{1/3} - r}{3 {\beta_0}^{1/3}} \right) }^2,
 \label{e44}
\end{equation}
\begin{equation}
 3 { (t+\beta_0) }^{1/3} = 3 {\beta_0}^{1/3} - r.
 \label{e45}
\end{equation}
Therefore, along the past light cone the solution (\ref{e25}) of the field
equation and its derivatives take the form
\begin{equation}
 R= \frac{r}{9} {(3 {\beta_0}^{1/3} - r) }^2; \ \ \ 
 R'=\frac{1}{9} {(3 {\beta_0}^{1/3} - r) }^2; \ \ \ 
 {\dot{R}}'=2 {(3 {\beta_0}^{1/3} - r) }^{-1}. \ \ \ 
 \label{e46}
\end{equation}
With the results above, we can easily obtain the observational relations
in the Einstein-de Sitter model. The redshift, luminosity distance,
cumulative number count, observed volume, volume and local densities are
respectively given by
\begin{equation}
 1+z= { \left( \frac{3 {\beta_0}^{1/3}}{3 {\beta_0}^{1/3} - r} \right) }^2,
 \label{z}
\end{equation}
\begin{equation}
 d_\ell=\frac{9 r {\beta_0}^{4/3}}{ { ( 3 {\beta_0}^{1/3} - r)}^2},
  \label{e47}
\end{equation}
\begin{equation}
  N_c= \frac{2r^3}{9M_G},
  \label{e48}
\end{equation}
\begin{equation}
  V=\frac{12 \pi r^3 {(3 \beta_0)}^4}{{( 3 {\beta_0}^{1/3} - r)}^6},
  \label{e49}
\end{equation}
\begin{equation}
  \rho_v=\frac{{( 3 {\beta_0}^{1/3} - r)}^6}{54 \pi {(3 \beta_0)}^4},
  \label{e50}
\end{equation}
\begin{equation}
  \rho=\frac{1}{6 \pi} { \left( \frac{3}{ 3 {\beta_0}^{1/3}-r} \right) }^6.
  \label{e51}
\end{equation}

We can see that the local and volume densities change along the past null
geodesic, and this is obvious from the dependence on $r$ in both equations
(\ref{e50}) and (\ref{e51}). That happens because along the past light cone
the integration goes through different surfaces of $t$ constant, and in this
sense {\it the Einstein-de Sitter model does appear to be inhomogeneous along
the backward null cone}. The presence of the coordinate $r$ in equation
(\ref{e51}) does not mean a spatially inhomogeneous local density, since $r$
is only a parameter along the past null geodesic, and due to equation
(\ref{e45}) each value of $r$ corresponds to a single value of $t$, that is,
each $r$ corresponds to a specific $t=$constant hypersurface given by
equation (\ref{e45}). However, as along the null geodesic the local density
effectively changes as it goes through different surfaces of constant time 
coordinate, hence in this sense the model can be thought of as
inhomogeneous. The volume density may also change inasmuch as it is being
measured through these same hypersurfaces of $t$ constant, where each one has
different  values for the proper density, and being a cumulative density,
$\rho_v$ averages at bigger and bigger volumes in a way that adds more and
more different local densities of each spatial section of the model.

Another interesting result follows if we look at the asymptotic limit of
the equations above. As the function $\beta(r)$ determines the local
time at which $R=0$ (see the appendix), the surface $t+ \beta=0$ is a
surface of singularity, which means that the physical region considered
is given by the condition $t+ \beta > 0$. Considering equation (\ref{e45}),
it is obvious that $r= 3 {\beta_0}^{1/3}$ corresponds to the surface of
singularity, meaning that the physical region of the model is given by
the condition $0 \le r \le 3 {\beta_0}^{1/3}$. Therefore, when $r \rightarrow
3 {\beta_0}^{1/3}$ the observational relations breakdown since $z \rightarrow 
\infty$, $V \rightarrow \infty$, $d_\ell \rightarrow \infty$, $\rho
\rightarrow \infty$ and $\rho_v \rightarrow 0$.

So we can see that the volume density vanishes asymptotically when
observables are plotted, and this result is a consequence of the
definition adopted here for the volume density since at the big bang
singularity hypersurface the observed volume is infinity, but the total 
mass is finite. Thus we may say that the following limit holds in the
Einstein-de Sitter cosmology:
\begin{equation}
 \lim_{d_\ell \rightarrow \infty} \rho_v = 0.
 \label{limite}
\end{equation}
Comparing with equation (\ref{e7}), this result means that this
relativistic cosmology obeys the zero global density postulate
for a pure hierarchy, as defined by Wertz (and conjectured later by
Pietronero), and therefore, we may that {\it under the appropriate
definitions the Einstein-de Sitter model seems to meet all postulates of
a pure hierarchical model, in the sense of Wertz}.

The zero global density postulate was formulated by Wertz since it is a
logical result if one takes the Newtonian version of
the de Vaucouleurs density power law to its logical asymptotic limit.
However, this result has been repeatedly used as a fundamental reason why
the universe cannot be hierarchical (or fractal) at larger scales because
this postulate not only supposedly contradicts the spatially homogeneous
Friedmann model \cite{ruffini-song-taraglio}, thought to be the correct
cosmological model, but also is considered conceptually unacceptable. For
instance, \cite{feng} states that ``if the universe is really the Friedmann
type on large scale (...) the inhomogeneous structure must cease on large
enough scale''. It is then clear from equation (\ref{limite}) that this
is may be a rather misleading approach to relativistic fractal cosmology,
and even to cosmology in general, since we
definitively can have an observationally based interpretation of the
Friedmann model where it has no well defined average density, and is
inhomogeneous, with asymptotically zero global density along the past
light cone. Therefore, it seems to be an inescapable conclusion that
having or not having zero global density in the model is just a question
of interpretation.~\footnote{ \ It is interesting to note that many
researchers are quick to reject any cosmology with a vanishing global
density, although, it could be argued, that the no lesser strange result
of an infinity local density at the big bang is accepted without argument.
On this respect it also could be argued that both results seem to be
exactly what they are, that is, at their face values they are just limits,
taken under different circumstances, where the observables breakdown.}

\begin{figure}[bht]
  \vspace{7.5cm}
  \epsfbox[0 0 30 50]{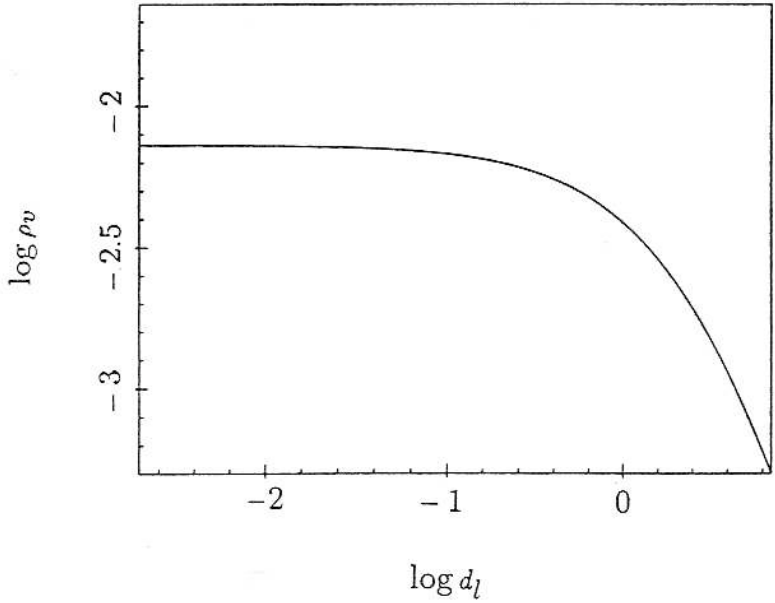}
  \caption{Plot of $\rho_v$ vs.\ $d_\ell$ in the Einstein-de Sitter
           model for the range $0.001 \le r \le 1.5$ and with $\beta_0=2.7$
           (distance is given in Gpc, and $H_0=75 \ \mbox{km}
           \ {\mbox{s}}^{-1} \ {\mbox{Mpc}}^{-1}$ is assumed). The 
           distribution does not appear to remain homogeneous along the
           null geodesic and the fractal dimension departs from the
           initial value 3 \protect\cite{paper2}.}\label{EdS}
\end{figure}

The inhomogeneity of the Einstein-de Sitter model can be graphically seen
in figure \ref{EdS} where the volume density is plotted against the luminosity
distance. At close ranges the model is homogeneous, with the average showing
no significant departure from a constant value. This means that at very
close ranges the volume density is being measured at our constant time
hypersurface, but once the scale increases the average starts to change,
which means that $\rho_v$ begins to be calculated at regions where the local
density differs from the value at our ``now'' ($t=0$). In other words,
once the volume density starts to change significantly we would have a
significant distancing from the initial time hypersurface, going towards
earlier epochs of the model. The plot then shows, in a quantitative way,
the scale where theoretically the Einstein-de Sitter model no longer is
observationally homogeneous, which in the case of the plot of figure
\ref{EdS} is at about $z \approx 0.04$, or $d_\ell \approx 160$ Mpc.
It is easy to show that a 30\% decrease in $\rho_v$ from the value at $t=0$
(now) happens at about $z \approx 0.1$, or $d_\ell \approx 500$ Mpc,
and therefore, this is approximately the
maximum range up to where the homogeneity of the Einstein-de Sitter model can be
observed. In other words, as this limit is obtained by solving the past null
geodesic equation and using the result in observational relations, we have
here a clear evidence that relativistic effects become important in cosmology
at very close ranges, and this offers us an observationally based methodology
that in principle allows us to ascertain quantitatively such ranges in different
scenarios.

Another interesting aspect that comes out of the analysis of figure \ref{EdS}
is the fact that the fractal dimension of the distribution has only the
value $D=3$ in the homogeneous, or flat region of the plot. Beyond this it
starts to decrease, effectively making the thinning rate in equation
(\ref{e37}) a function of position which, in other words, is incompatible
with a single fractal description for the distribution of dust in the
Einstein-de Sitter cosmology.

\begin{figure}[bht]
  \vspace{8cm}
  \epsfbox[0 0 30 50]{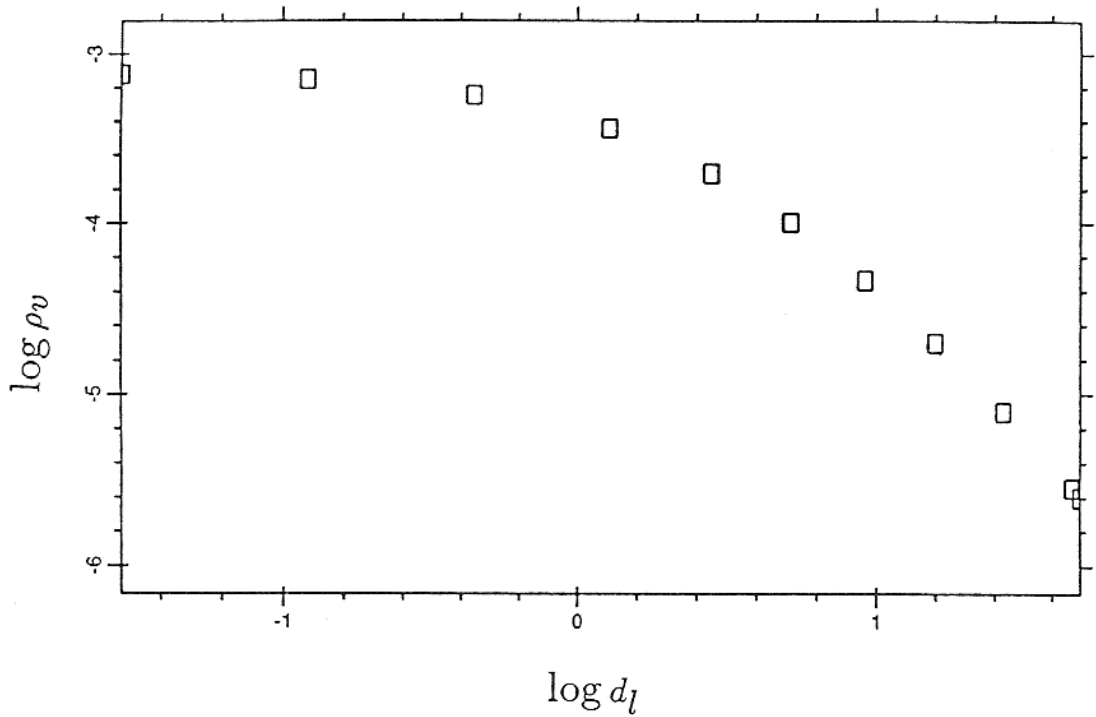}
  \caption{Numerical results for $\rho_v$ vs.\ $d_\ell$ in the open
           Friedmann model model for the range $0.001 \le r \le 1.5$
           and with $\beta_0=3.6$ and $\Omega_0 \approx 0.2$. The 
           deviation from a homogeneous initial region is also visible,
           and starts at about $d_\ell \approx 250$ Mpc
           \protect\cite{paper3}.}\label{open}
\end{figure}

The open and recollapsing Friedmann models also behave in a similar
inhomogeneous manner as compared to the Einstein-de Sitter model, although
the ranges where the departure from homogeneity starts are different.
Figure \ref{open} shows the numerical solution for the observational
\begin{figure}[bht]
  \vspace{8cm}
  \epsfbox[0 0 30 50]{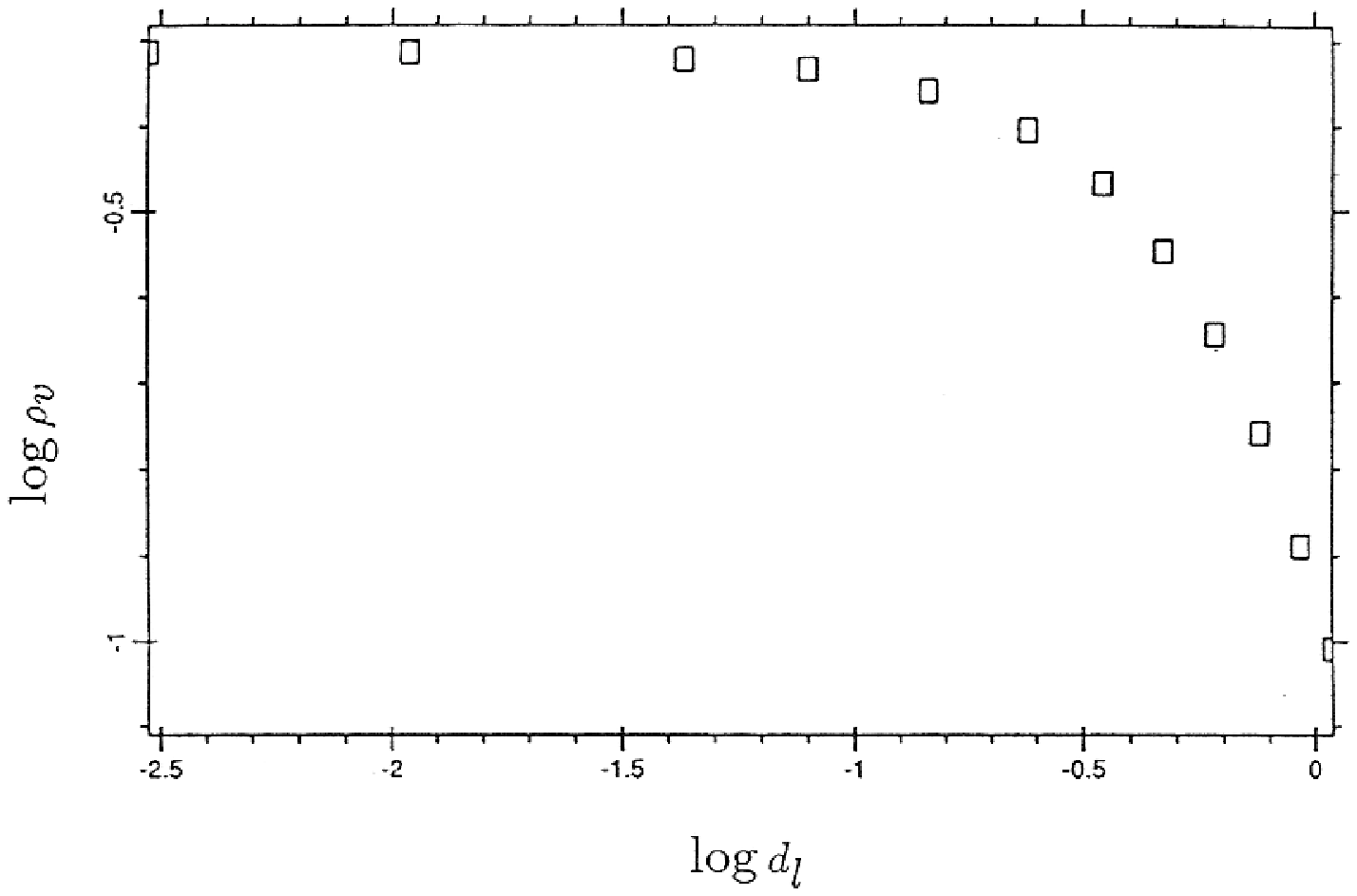}
  \caption{Numerical results for $\rho_v$ vs.\ $d_\ell$ in the
           recollapsing Friedmann model model for the range
           $0.001 \le r \le 1.5$ and with $\beta_0=0.7$ and
           $\Omega_0 \approx 4$. The departure from the homogeneous
           initial region starts at very close range, at about
           $d_\ell \approx 30$ Mpc \protect\cite{paper3}.}\label{close}
\end{figure}
relations in the $K=-1$ Friedmann model, and figure \ref{close} shows the
same plot for the recollapsing model \cite{paper3}. We can clearly see the
deviation from homogeneity in both models, although the former starts to
deviate at about $d_\ell \approx 250$ Mpc while in the latter that happens
at $d_\ell \approx 30$ Mpc. Due to the similarity of the graphs, it seems
reasonable to conclude that all Friedmann models appear to obey Wertz's
zero global density postulate. 

\newpage
\section{TOLMAN FRACTAL SOLUTIONS}

\indent

Once the observational relations and the numerical methodology for
finding fractal solutions in the Tolman model is developed, we can go to
the stage of actually specializing the free functions of Tolman's metric
in order to see which ones, if any, do have fractal behaviour.~\footnote{ \ The
numerical code written in {\sc fortran 77} used to solve the past
null geodesic in the Tolman metric and to find fractal solutions is published
in \protect\cite{thesis}. [{\it Note added in 2009-- See also Ref.\ 
\cite{rib2002} and} {\tt http://www.if.ufrj.br/$\sim$mbr/codes/}.]}
Nonetheless, some criteria must be met by those
solutions such that some essential observational constraints are obeyed.
Those criteria for choosing and accepting the solutions can be listed as
follows:
\begin{itemize}
 \item linearity of the redshift-distance relation for $z < 1$;
 \item the Hubble constant within the currently accepted range
       $40 \ \mbox{km} \ {\mbox{s}}^{-1} \ {\mbox{Mpc}}^{-1} < H_0 <
       100 \ \mbox{km} \ {\mbox{s}}^{-1} \ {\mbox{Mpc}}^{-1}$;
 \item constraint in the fractal dimension of $1 < D < 2$;
 \item obedience to the Vaucouleurs' density power law $\rho_v \propto
       {d_\ell}^{- \gamma}$.
\end{itemize}
By means of the methodology described in section 4, a systematic search
for fractal solutions was carried out \cite{paper3}, but only hyperbolic
type solutions were found to meet all the criteria outlined above. Here
I shall only show the simplest hyperbolic fractal solution obtained, since
the other classes of fractal solutions did not meet all criteria above.
A complete discussion and presentation of the other solutions can be found
in \cite{paper3}.

\begin{figure}[bht]
  \vspace{8cm}
  \epsfbox[0 0 30 50]{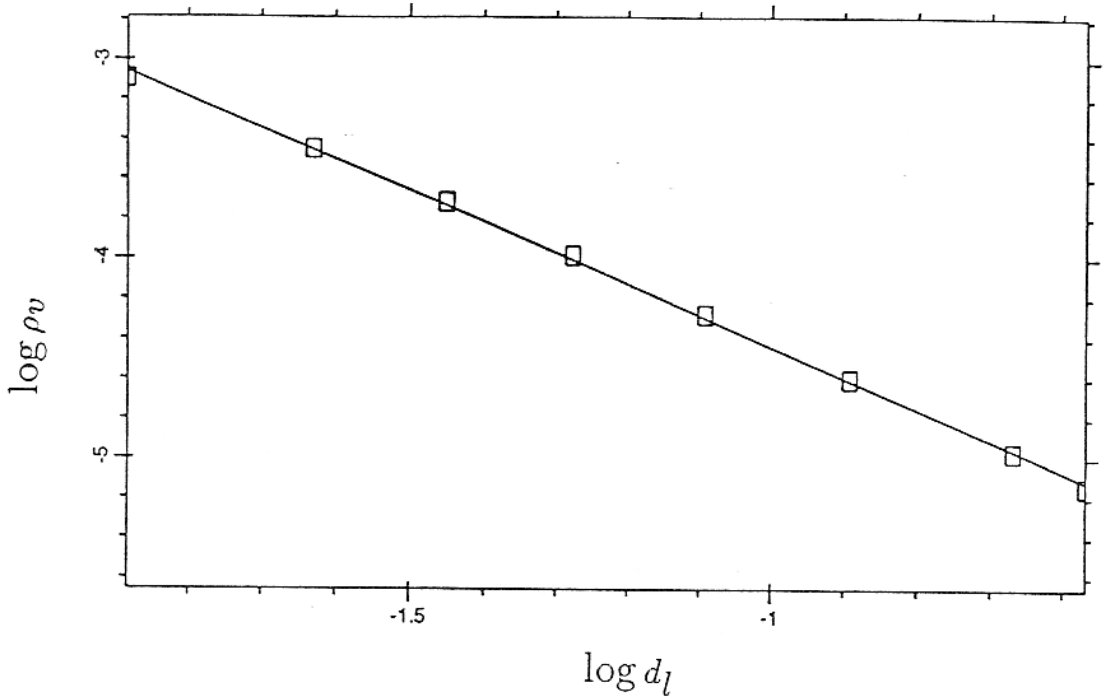}
 \caption{Results of the volume density $\rho_v$ vs.
      luminosity distance $d_l$ of the hyperbolic model (\protect\ref{t30}).
      The integration is in the interval $0 \leq
      r \leq 0.07$ and the constants are $\alpha=10^{-4}, \ p=1.4, \
      \beta_0=3.6$. The fitting coefficients calculated are $a_1=-6.0$
      and $a_2=-1.6$, giving a fractal dimension $D=1.4$ and a lower
      cutoff constant $\sigma=8.7 \times 10^5$.}\label{f1}
\end{figure}

\begin{figure}[bht]
  \vspace{9cm}
  \epsfbox[0 0 30 50]{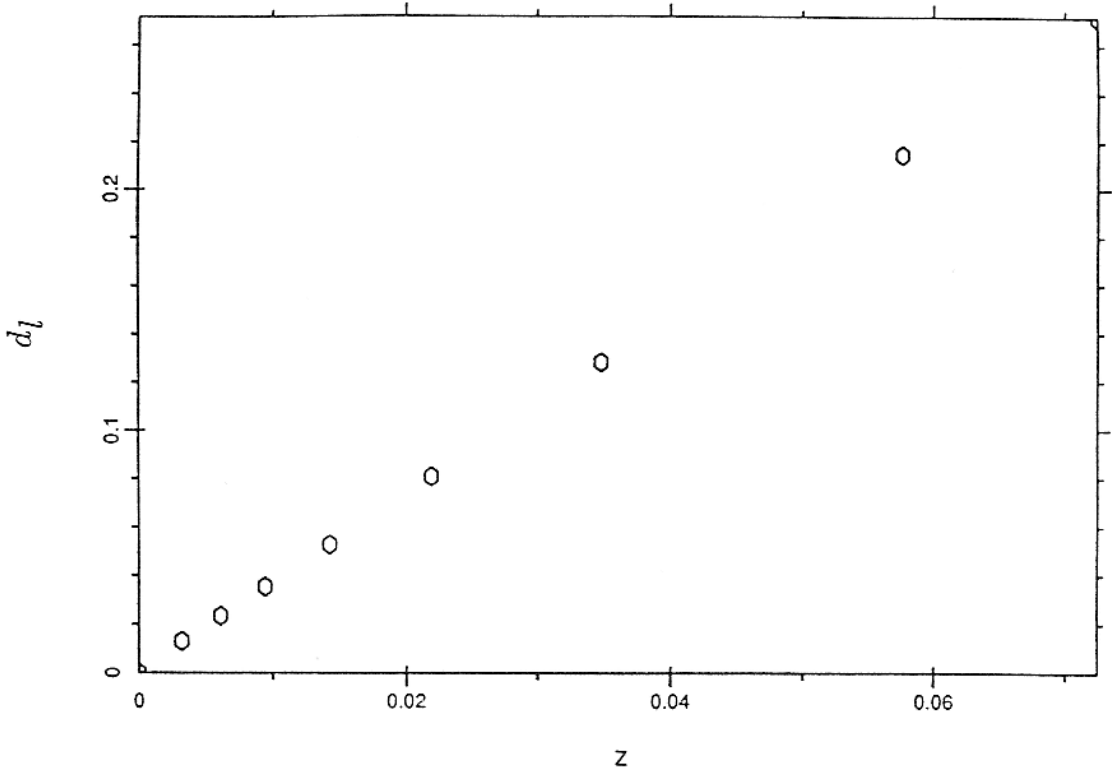}
 \caption{Distance-redshift diagram obtained with the
      hyperbolic model of functions (\protect\ref{t30}). The slope of the
      diagram obtained for $d_\ell$ vs.\ $z$ gives $H_0 \protect\cong 80 \ 
      \mbox{km} \ {\mbox{s}}^{-1} \ {\mbox{Mpc}}^{-1}$, a
      value which is within the current uncertainty in the Hubble
      constant.}\label{f2}
\end{figure}

\begin{figure}[bht]
  \vspace{7.5cm}
 \epsfbox[0 0 30 50]{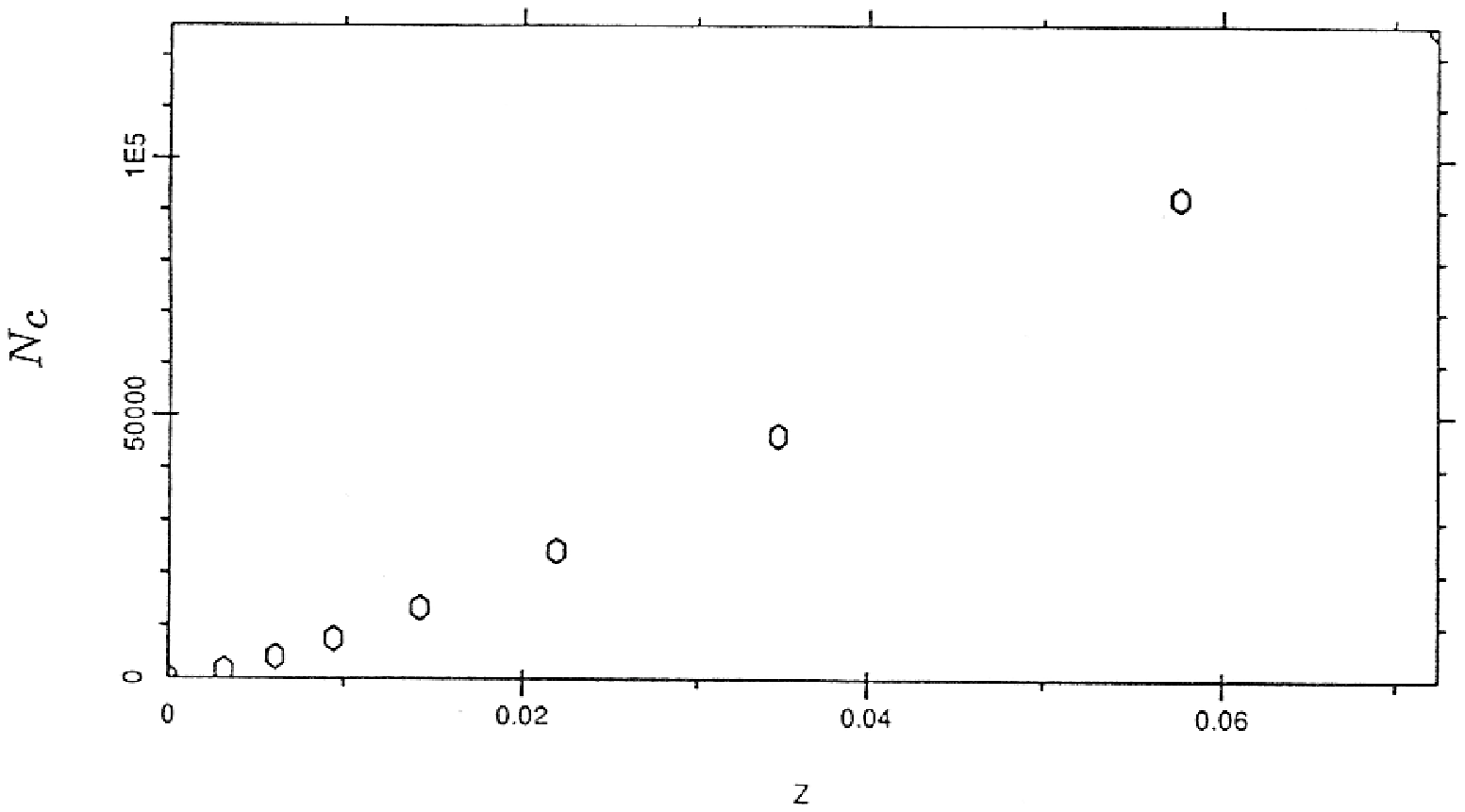} 
 \caption{Plot of the results for the cumulative number
      counting $N_c$ vs.\ the redshift $z$ given by the integration of
      the model (\protect\ref{t30}) with the same parameters as in figure
      \protect\ref{f1}.}\label{f3}
\end{figure}

The particular form of the three functions that led to fractal behaviour
in hyperbolic models is as follows:
\begin{equation}
 \left\{ \begin{array}{ll}
           f = \cosh r, \\
           F = \alpha r^p, \\
           \beta = \beta_0,
         \end{array}
 \right.
 \label{t30}
\end{equation}
where $\alpha$, $p$, and $\beta_0$ are positive constants. The
experience with the numerical simulations show that $\alpha$ must be
around $10^{-4}$ to $10^{-5}$ and $p$ and $\beta_0$ can vary from around
0.5 to 4. Figure \ref{f1} shows the power law behaviour of $\rho_v$ vs.
$d_\ell$ of the model formed by functions (\ref{t30}), with a fractal
dimension of $D=1.4$. The straight line fitted according to equation
(\ref{e40}) is also clearly visible. Figure \ref{f2} is the
redshift-distance diagram of the same model where we can see the good
linear approximation given by functions (\ref{t30}). The slope of the
points gives $H_0 \cong 80 \ \mbox{km} \ {\mbox{s}}^{-1} \ {\mbox{Mpc}}^{-1}$,
and it is interesting to note that recent measurements made by two different
methods suggest a Hubble constant very close to this value \cite{peacock}.
Actually, for $\beta_0 = 3.6$, the value used to get the results shown
in figures \ref{f1} and \ref{f2}, we would have an age of the universe
of about 12 Gyr, which is a lower limit if we consider the age of
globular clusters \cite{peacock}. Therefore, also in this point of the
age of the universe, the model (\ref{t30}) agrees reasonably well with
observations. The integrations with functions (\ref{t30}) were stopped
at $z \cong 0.07$, which corresponds to the luminosity distance
$d_\ell \cong 270$ Mpc and this is the redshift depth of the {\sc iras}
redshift survey \cite{saunders}. Finally, figure \ref{f3} shows the
results for cumulative number counting vs.\ redshift produced by the
model under consideration.

As mentioned above, there are strict limits on the values of the
parameters $\alpha$, $p$ and $\beta_0$, and by means of numerical
experimentation it was found that outside the ranges stated above, the
fractality of the model (\ref{t30}) is destroyed. Therefore, it seems
that this fractal model is {\it structurally fragile} as a variation of
the parameters of the model can produce a qualitative change in its
behaviour.

As closing remarks, it can be shown \cite{paper3} that the model (\ref{t30})
remains fractal at different epochs, with a remarkable constancy in the
fractal dimension. In addition, if a Friedmann metric is joined to the
solution (\ref{t30}), we can use the internal Tolman solution to find
the Friedmann model which best fits another cosmological model that
gives a realistic representation of the universe (in this case, the
Tolman model given by equations [\ref{t30}]), including all inhomogeneities
down to some specified length scale mandatory. By doing this procedure,
it can be shown \cite{paper3} that the best Friedmann model is an open
one, with $H_0 = 83 \ \mbox{km} \ {\mbox{s}}^{-1} \ {\mbox{Mpc}}^{-1}$ and
$\Omega_0 \cong 0.002$. This low value for $\Omega_0$ is explained as
due to the fact that in this approach no kind of dark matter was
considered, but only the luminous matter associated with the galaxies,
which are assumed to form a fractal system. Galactic luminous matter
gives a value for $\Omega_0$ of the same order of magnitude as the one
found above.

Joining an external Friedmann metric to the internal inhomogeneous
solution was initially thought to be a good way of modelling a possible
crossover to homogeneity to the fractal system. However, as the Friedmann
metric looks inhomogeneous at larger scales when measured along the past
light cone, this result seems to imply that this external solution is not
really mandatory.

\section{CONCLUSIONS AND DISCUSSION}

\indent

This work presented in essence a different approach for modelling the
large scale distribution of galaxies, whose main idea is to assume the
non-orthodox, but old principle that the empirically observed
self-similarity of this distribution is a fundamental fact to be taken
into account in any model that attempts a realistic representation of
the distribution of galaxies. Under this philosophy, a simple relativistic
model was advanced, model which is essentially a translation to a
relativistic framework of the Newtonian hierarchical (fractal) cosmology
developed by Wertz and Pietronero. This is a simple exploratory model,
which although it had to assume some strong simplifications, it is able
to obtain some interesting new results, like the inhomogeneity of the
Friedmann model, once the observational relations derived for this fractal
model are used in this metric. It also shows that the idea of a vanishing
global density, postulated for hierarchical (fractal) cosmologies, is not
necessarily in contradiction with currently accepted cosmological models
like the Einstein-de Sitter one, being essentially a question of using the
appropriate definitions for density and distance. A general methodology
for searching fractal solutions is also advanced, and this methodology
is able to successfully find them, showing that solutions of Einstein's
field equations approximating a single fractal structure do exist.

In addition to those points, some remarks about the method and the solution
should be made. In the first place, the best numerical simulation
presented an open, ever expanding model for the large scale distribution
of galaxies, and although such class of models are obviously favoured by
the simulations, it must be said that flat or recollapsing models are not
at all ruled out as there may be some more complex forms for the functions
$f(r)$, $F(r)$ and $\beta(r)$ which produce observationally compatible
models, but which were not investigated. Secondly, in view of the results
obtained with the fractal inspired observational relations, it seems that
observable average densities appear to be physically interesting for the
characterization of cosmological models. In particular they are important
in the description of a fractal system, and in the determination of the
limits of validity of the homogeneous hypothesis. Thirdly, as discussed
in the Introduction, the model showed here provides a {\it description} for
the possible fractal structure of the universe, but as in the problem of
coastlines, it does not provide an answer to the question of where this
fractal system came from, its origins, and why the large scale luminous
matter appears to follow a fractal pattern.

To try to answer this last point, it may be crucial a deeper study of the
dynamics of the field equations, or even the particular model showed here,
as their nonlinearity may provide some important clues such that one may be
able to get closer to answering the question of where this fractal pattern
came from. In this respect there are some aspects that may possibly help in
this direction. In the first place, structural fragility seems to be a
reality in the fractal model discussed (and the others presented in 
\cite{paper3}), since there is evidence for this behaviour from the
numerical experimentations themselves. Secondly, from the theory of
dynamical systems we know that strange attractors have a fractal pattern
in phase space. Therefore, the obvious questions is when we see, along
the past null geodesic, a self-similar fractal pattern, could that mean
that a strange attractor is lurking from behind the scenes? In this case,
how can we characterize it? In addition to this, as in some chaotic
dynamical systems, could there exists a possible link between the fractal
dimension and the Lyapunov exponents in this cosmological context?

Those questions are already in the realm of the second stage of the 
physicist's approach to dynamical systems, as discussed in the
Introduction, and by their own nature, answering them will
probably be a very challenging task. In any case, a complete or partial
response to any of these points is likely to shed some light on the
underlying dynamics of the relativistic field equations in this specific
model or, perhaps, in more general ones.

%%%%%%%%%%%%%%%%%%%%%%%%%%%%%%%%%%%%%%%%%%%%%%%%%%% acknowledgements

\section*{ACKNOWLEDGMENTS}

\indent

Thanks go to the organizing committee for this interesting and
informative workshop, and in special to David Hobill for the smooth way
the local organization was carried out.

%%%%%%%%%%%%%%%%%%%%%%%%%%%%%%%%%%%%%%%%%%%%%%%%%%% references
%\newpage

%%%%%%%%%%%%%%%%%%%%%%%%%%%%%%%%%%%%%%%%%%%%%%%%%%% appendix
\appendix
\section*{APPENDIX. The Friedmann metric as a special case of the
		    Tolman solution}

\indent

The aim of this appendix is to show how the Tolman solution
can be reduced to the Friedmann metric by means of the specializations of
the functions $f(r)$, $F(r)$, $\beta(r)$, and the physical differences
between the two metrics. This may be useful for those not familiar with
the Tolman solution.

It can be shown, by calculating the junction conditions between the
Tolman and Friedmann metrics \cite{paper1}, that in order to obtain the
latter from the former we have to assume that 
   \begin{equation}
    R(r,t) = a(t) \ g(r), \ \ \ \ f(r) = g'(r).
    \label{A1}
  \end{equation}
  By substituting equations (\ref{A1}) into the metric (\ref{e22}) we get
  \begin{equation}
   dS^2=dt^2-a^2(t) \left\{ dr^2 + g^2(r) \left[ d \theta^2+\sin^2 \theta
	d \phi^2 \right] \right\},
   \label{A2}
  \end{equation}
  which is a Friedmann metric if
  \begin{equation}
    g(r) = \left\{ \begin{array}{ll}
	           \sin r,     &   \\
		      r,       &   \\
	           \sinh r.    &   
	           \end{array}
           \right.
     \label{A3}
  \end{equation}
  If we now substitute equations (\ref{A1}) in equation (\ref{e24}) and
  integrate it, that gives
  \begin{equation}
   \frac{F}{4} = \frac{4 \pi}{3} \rho a^3 g^3.
   \label{A4}
   \end{equation}
   It is worth noting that the time derivative of the equation
   above gives the well-known relation for the matter-dominated era of
   a Friedmann universe:
   \[
     \frac{d}{dt} \left( \rho a^3 \right) = 0.
   \]
   Equation (\ref{A4}) is necessary in order to deduce the usual Friedmann
   equation. This is possible by substituting equations (\ref{A1}) and
   (\ref{A4}) into equation (\ref{e23}). The result may be written as
   \begin{equation}
    \dot{a}^2 = \frac{8 \pi}{3} \rho a^2 - K 
   \label{A5}
   \end{equation}
   where
   \[
     K = \frac{1-{g'}^2}{g^2}.
   \]
   It is easy to see that $K=+1, \ 0, \ -1$ if $g=\sin r, \ r, \ \sinh
   r$, respectively, and this shows that equation (\ref{A5}) is indeed the
   usual Friedmann equation. Let us now write equation (\ref{A5}) in the
   form
   \begin{equation} 
       \frac{\dot{a}^2 g^2}{2} - \frac{m}{ag} = - \left(1-{g'}^2 \right)
    \label{A7}
   \end{equation} 
   where
   \begin{equation}
    m(r) = \frac{4 \pi}{3} \rho a^3 g^3.
    \label{A8}
   \end{equation}
   Equation (\ref{A7}) is interpreted as an energy equation
   \cite{bondi} and, in consequence, $m(r)$ is the gravitational
   mass inside the coordinate $r$. Thus $4 m(r) = F(r)$ and this shows the
   role of the function $F(r)$ in providing the gravitational mass of the
   system. In addition, equation (\ref{A7}) shows that the function $f(r)$
   in Tolman's spacetime gives the total energy of the system.

   The function $\beta(r)$ gives the big bang time of the model, and this
   can be seen as follows. If ``now'' is defined as $t=0$ and if
   $\beta(r) = 0$, then the hypersurface $t=0$ is singular, that is,
   $R=0$ everywhere \footnote{ \ This is also valid for $f^2 > 1$ and $f^2
   < 1$ type solutions of equation (\ref{e23}).}. So $\beta(r)$ 
   gives the age of the universe which
   in Tolman's spacetime may change if different observers are situated
   at different radial coordinates $r$. This is a remarkable departure
   from Friedmann's model that gives the same age of the universe for
   all observers on a hypersurface of constant $t$. In other words, in
   a Friedmann universe the big bang is simultaneous while in a Tolman one
   it may be non-simultaneous, that is, the big bang may have occurred
   at different proper times in different locations. As a consequence 
   another essential ingredient in reducing the Tolman metric to
   Friedmann is $\beta =$ constant, and so the linkage
   between $\beta$ and the Hubble constant is of the form
   \begin{equation}
    \frac{\dot{R}}{R} = \frac{\dot{a}}{a} = H(t), \ \ \ \mbox{for} \ \ 
    \beta = \beta_0,
    \label{A9}
   \end{equation}
   where $\beta_0$ is a constant. Considering equation (\ref{e25}) it is
   straightforward to conclude that
   \begin{equation}
    \beta_0 = \frac{2}{3 H_0},
   \label{A10}
   \end{equation}
   where $H_0 = H(0)$. Equation (\ref{A10}) gives the relationship 
   between $\beta_0$ and the Hubble constant $H_0$ in a Einstein-de
   Sitter universe.

Bearing this discussion in mind it is easy to see that the usual
Friedmann universe requires that $g= \sin r$, $r$, $\sinh r$ and
$F= b_1 \sin^3 r$, $\frac{8}{9} r^3$, $ b_2 \sinh^3 r$,
which are respectively the cases for $K=+1$, $0$, $-1$. The positive
constants $b_1$ and $b_2$ are scaling factors necessary to make the
density parameter $\Omega$ equal to any value different from one in the
open and closed models.

Let us now extend equation (\ref{A10}) to the $K= \pm 1$ Friedmann cases.
Considering the specializations above, that permit us to get the Friedmann
metric from the Tolman solution, and assuming $t=0$ as our ``now'', that
is, $t=0$ being the time coordinate label for the present epoch, it is
straightforward to show that the present value $H_0$ for the Hubble
constant in the closed Friedmann model is given by
\begin{equation}
 H_0 = \frac{4 \sin 2 \Theta_0}{b_1 {(1- \cos 2 \Theta_0 )}^2},
  \label{add-1}
\end{equation}
where $\Theta_0$ is the solution of
\begin{equation}
   4 \beta_0 = b_1 (2 \Theta_0 - \sin 2 \Theta_0),
    \label{add-2}
\end{equation}
and $\beta(r)=\beta_0$ is a constant that gives the age of the universe. 
In the open Friedmann case we will have then
\begin{equation}
     H_0 = \frac{4 \sinh 2 \Theta_0}{b_2 {(\cosh 2 \Theta_0 - 1 )}^2},
      \label{add-3}
\end{equation}
and
\begin{equation}
  4 \beta_0 = b_2 ( \sinh 2 \Theta_0 - 2 \Theta_0).
  \label{add-4}
\end{equation}

For the sake of completeness, let us now obtain the value of the
cosmological density parameter $\Omega$ in the Friedmann model at the
present constant time hypersurface. By definition $\Omega~=~\rho~/~\rho_c$
and $\rho_c~=~1/~(6~\pi~{\beta_0}^2)$ at $t=0$. Therefore, in the closed
Friedmann model we have that
\begin{equation}
    \Omega_0 = \frac{72 {\beta_0}^2}{ {(b_1)}^2 {(1- \cos 2
      \Theta_0)}^3},
  \label{add-5}
\end{equation}
where $\Theta_0$ is given by equation (\ref{add-2}). In the open 
Friedmann model we have
\begin{equation}
  \Omega_0 = \frac{72 {\beta_0}^2}{ {(b_2)}^2 {(\cosh 2
  \Theta_0 - 1)}^3},
    \label{add-6}
\end{equation}
with $\Theta_0$ being the solution of equation (\ref{add-4}).

\end{document}